\documentclass[twocolumn,showpacs,preprintnumbers,amsmath,amssymb]{revtex4}

\usepackage{amsmath,amssymb,graphicx}
\usepackage{graphicx}
\usepackage{dcolumn}
\usepackage{bm}
\usepackage{epsfig}
\usepackage{here}
\usepackage{float}
\usepackage{amsmath}
\usepackage[usenames]{color}

\newcommand{\bea}{\begin{eqnarray}}
\newcommand{\eea}{\end{eqnarray}}

\begin{document}

\title{Perturbation initial conditions for a couple of dark energy scalar field potentials}
\author{Jai-chan Hwang${}^{1}$, Hyerim Noh${}^{2}$, Chan-Gyung Park${}^{3}$, Da-Hee Lee${}^{1}$}
\address{
         ${}^{1}$Department of Astronomy and Atmospheric Sciences, Kyungpook National University, Daegu, Korea \\
         ${}^{2}$Center for Large Telescope,
         Korea Astronomy and Space Science Institute, Daejon, Korea \\
         ${}^{3}$Division of Science Education and Institute of Fusion Science, Chonbuk National University, Jeonju, Korea
         }


\begin{abstract}
We present perturbation initial conditions for two types of scalar field potential often used in the dark energy study: one inverse power-law and the other exponential. The solutions are presented in the presence of the $w =$ constant fluid ($w = 1/3$ for radiation fluid), a minimally coupled scalar field and a sub-dominating zero-pressure fluid (cold dark matter and baryon dust). We consider two gauge conditions, the $w$-fluid comoving gauge and the cold-dark-matter comoving gauge; solutions in the latter gauge are derived by the gauge transformation and this method can be applied to derive solutions in any other gauge condition.
\end{abstract}

\noindent 

\maketitle

\section{Introduction}

The minimally coupled scalar field (MSF) is widely used as a model of dark energy dynamically providing the late time cosmic acceleration \cite{DE-reviews}. In the literature the MSF is often replaced by an effective fluid with entropic perturbation. In order to handle the MSF directly as the dark energy we need the proper initial conditions (in the early radiation dominated stage and in the large-scale limit) for the perturbation as well as a background. Background solutions for a couple of field potentials are known in the literature. Based on background solutions, in this work we derive the perturbation initial conditions for the fluid and field system.

Here we consider three-component system: an ideal fluid with constant $w \equiv p_w/\mu_w$, a minimally coupled scalar field (MSF) $\phi$, and a sub-dominating zero-pressure fluid; $p_I$ and $\mu_I$ are the pressure and energy density with $I = w, c$ and $\phi$ for the $w$-fluid, cold dark matter (CDM) and the scalar field, respectively; baryon as a pressureless fluid can be absorbed to the CDM component. We consider two types of scalar field potential with the known background evolution: these are the inverse power-law potential with $V = \kappa \phi^{-\alpha}$ and the exponential potential with $V = V_0 e^{\lambda \phi}$. These potentials in fact cover wide range of more complicated potentials often used as the dark energy in the literature in the early evolution era.

We present the perturbation initial conditions for these two types of potential in the two gauge conditions: the $w$-fluid comoving gauge ($w$CG) and the CDM comoving gauge (CCG). Solutions for the exponential potential were presented previously but only in the two-component system in the $w$CG \cite{HN-2001-scaling}. We derive solutions in the CCG from the ones in the $w$CG using the gauge transformation. This shows how one could apply the same procedure to get solutions in any other gauge condition. The solutions in the CCG are particularly relevant in research as the state of the art publicly available CAMB code is based on this gauge \cite{CAMB}.

A complete set of background and perturbation equations with multiple fluids and a field is summarized in the Appendix. We set $c \equiv 1 \equiv \hbar$.

\section{Inverse power-law potential}

\subsection{Background solutions}

We consider an inverse power-law potential \cite{Peebles-Ratra-1988}
\bea
   & & V = {\kappa \over \phi^\alpha},
   \label{inv-potential}
\eea
with constant $\alpha$ and $\kappa$.
In the {\it $w$-fluid dominated} era, thus $\mu_\phi / \mu_w \ll 1$, we have
\bea
   H^2 = {8 \pi G \over 3} \mu_w, \quad
       \mu_w \equiv \mu_{w0} a^{-3(1+w)}, \quad
       a \propto t^{2 \over 3(1+w)},
\eea
where $H \equiv \dot a/a$ with $a(t)$ the cosmic scale factor, and we set $a_0 \equiv 1$ with the subscript $0$ indicating the present epoch. The equation of motion becomes
\bea
   & & \phi^{\prime\prime}
       + {3 (1 - w) \over 2} \phi^{\prime}
       + {1 \over H^2} V_{,\phi} = 0,
   \label{EOM-0}
\eea
where a prime indicates the derivative in terms of $x \equiv \ln{a}$.

By setting $\phi \equiv C a^n = C e^{nx}$ we find a solution \cite{Peebles-Ratra-1988,Watson-Scherrer-2003}
\bea
   & & n = {3 (1 + w) \over \alpha +2},
   \nonumber \\
   & &
       C = \left( {\kappa \alpha (\alpha + 2)^2 \over
       12 \pi G \mu_{w0} (1 + w) (\alpha + 4 - \alpha w)}
       \right)^{1/(\alpha+2)}.
   \label{BG-IC-IPL-1}
\eea
We have
\bea
   & & {\mu_\phi \over \mu_w}
       = {1 + w \over \alpha (\alpha + 2)} 24 \pi G \phi^2,
   \nonumber \\
   & &
       V = {\alpha + 4 - \alpha w \over 2(\alpha+2)} \mu_\phi, \quad
       {\dot \phi^2 \over 2}
       = {\alpha (1 + w) \over 2 (\alpha + 2)} \mu_\phi.
   \label{BG-IC-IPL-2}
\eea
These solutions remain valid in the presence of additional sub-dominating fluid, like a zero-pressure fluid.

\subsection{Solutions in the $w$-fluid comoving gauge ($w$CG)}

The $w$CG sets $v_w \equiv 0$ as the slicing condition. In the {\it $w$-fluid dominated} era, thus ignoring $\mu_\phi/\mu_w$  and $\mu_c/\mu_w$ order terms, from Eqs.\ (\ref{eq4}), (\ref{eq6-I}), (\ref{eq7-I}) and (\ref{EOM}) we have
\bea
   & & \delta_w^{\prime\prime}
       + {1 - 9 w \over 2} \delta_w^\prime
       - \left[ w {\Delta \over a^2 H^2}
       + {3 \over 2} (1 + 3 w) (1 - w) \right] \delta_w
   \nonumber \\
   & & \qquad
       = {6 (1 + w)^2 \over \alpha +2} 8 \pi G \phi \delta \phi^\prime
   \nonumber \\
   & & \qquad
       + {9 (\alpha + 4 - \alpha w) (1 + w)^2 \over 2(\alpha + 2 )^2}
       8 \pi G \phi \delta \phi,
   \label{delta-phi-eq-1} \\
   & & \delta \phi^{\prime\prime}
       + {3 \over 2} (1 - w) \delta \phi^\prime
   \nonumber \\
   & & \qquad
       + \left[ - {\Delta \over a^2 H^2}
       + {9 (\alpha + 1) (\alpha + 4 - \alpha w) (1 + w) \over
       2 (\alpha + 2)^2} \right] \delta \phi
   \nonumber \\
   & & \qquad
       = {3(1 - w) \over \alpha + 2} \phi \delta_w^\prime
       - {9 (\alpha + 4 - \alpha w) w \over (\alpha + 2)^2}
       \phi \delta_w.
   \label{delta-phi-eq-2}
\eea
In the {\it large-scale limit} we ignore $\Delta/(aH)^2$ order terms. By setting $\delta_w \propto e^{mx}$ and $\delta\phi \propto e^{\ell x}$, we find
\bea
   & & \left[ m - 1 - 3 w \right]
       \left[ m + {3 \over 2} \left( 1 - w \right) \right]
   \nonumber \\
   & & \qquad \times
       \bigg[ m^2
       - {3 \over 2} {2 - \alpha + 6 w + \alpha w \over 2 + \alpha} m
   \nonumber \\
   & & \qquad
       + {9 \over 2} {(1 + w) ( 2 + \alpha + 2 w - \alpha w) \over 2 + \alpha} \bigg]
       = 0,
   \nonumber \\
   & & \left[ \ell - \frac{5+\alpha+9w+3\alpha w}{\alpha+2} \right]
       \left[ \ell + \frac{3(\alpha-4w-\alpha w)}{2(\alpha+2)} \right]
   \nonumber \\
   & & \qquad \times
       \bigg[ \ell^2 + {3 \over 2} (1 - w) \ell
   \nonumber \\
   & & \qquad
       + {9 \over 2} {(\alpha + 1) (\alpha + 4 - \alpha w) (1 + w) \over
       (\alpha + 2)^2} \bigg]
       = 0,
\eea
with two real solutions. For the growing mode, we have
\bea
   & & m = 1 + 3 w, \quad
       \ell = \frac{(1 + 3w) \alpha + 5 + 9w}{\alpha+2},
\eea
and
\bea
   & &
       \delta \phi = \frac{3[(1 - w) \alpha + 2 - 8w - 6w^2 ]}{(\alpha+2)[(7 + 9w) \alpha + 29 + 60w + 27w^2 ]}
       \phi \delta_w,
   \nonumber \\
   & & \delta_w \propto a^{1 + 3 w}.
   \label{sol-inv-wCG-1}
\eea

In the presence of additional but subdominant CDM component we recover the same equations in (\ref{delta-phi-eq-1}) and (\ref{delta-phi-eq-2}). Thus, the presence of CDM does not affect the $w$-$\phi$ system. For the CDM-component, from Eqs.\ (\ref{eq6-I}) and (\ref{eq7-I}) we additionally have
\bea
   & & \delta_c^\prime
       = {3 w \over 1 + w} \delta_w
       + \overline \kappa
       + {\Delta \over a^2 H^2} \overline v_c,
   \label{eq6-c-wCG} \\
   & & \overline v_c^\prime + {3 (1 + w) \over 2} \overline v_c
       = - {w \over 1 + w} \delta_w,
   \label{eq7-c-wCG}
\eea
where we introduced dimensionless perturbation variables
\bea
   & & \overline v_c \equiv a H v_c, \quad
       \overline \kappa \equiv {\kappa_{\rm pert} \over H}, \quad
       \overline \chi \equiv H \chi.
\eea
[In order to distinguish $\alpha$ and $\kappa$ in Eq.\ (\ref{inv-potential}) from the perturbation variables used in the Appendix, we set the perturbed one as $\alpha_{\rm pert}$ and $\kappa_{\rm pert}$.] The variables $\delta_c$ and $v_c$ can be determined from solutions of $\delta_w$. In the $w$-fluid dominated era and in the large-scale limit, for the growing mode in Eq.\ (\ref{sol-inv-wCG-1}) the solutions are
\bea
   & & \delta_c = {1 \over 1 + w} \delta_w, \quad
       \overline v_c
       = - {2 w \over (1 + w) (5 + 9w)} \delta_w.
   \label{sol-inv-wCG-2}
\eea
From Eqs.\ (\ref{eq7-I}) and (\ref{eq6-I}), (\ref{eq2}) and (\ref{eq3}), respectively, we have
\bea
   & & \alpha_{\rm pert} = - {w \over 1 + w} \delta_w, \quad
       {\Delta \over a^2 H^2} \varphi
       = - {5 + 3 w \over 2 (1 + w)} \delta_w,
   \nonumber \\
   & &
       \overline \kappa = {1 \over 1 + w} \delta_w, \quad
       {\Delta \over a^2 H^2} \overline \chi
       = - {1 \over 1 + w} \delta_w.
   \label{sol-inv-wCG-3}
\eea
We can show that Eq.\ (\ref{eq5}) is satisfied, and Eq.\ (\ref{eq1}) gives
\bea
   & & \varphi^\prime
       = - {w \over 1 + w} \delta_w,
\eea
which is valid to next order in the large-scale expansion;
$\dot \varphi = 0$ for the mode in Eq.\ (\ref{sol-inv-wCG-3}).
We can show
\bea
   \delta_\phi
       =
       { \alpha [ (-1 + 24 w + 9w^2) \alpha
       - 2 + 84 w + 54w^2 ]
       \over
       2 (\alpha + 2)
       [ (7 + 9w) \alpha
       + 29 + 60 w + 27 w^2 ]} \delta_w.
   \label{sol-inv-wCG-5}
\eea
Equations (\ref{sol-inv-wCG-1}), (\ref{sol-inv-wCG-2})-(\ref{sol-inv-wCG-5}) are the complete solutions for inverse power-law potential in the $w$CG.

\subsection{Solutions in the CDM comoving gauge (CCG)}

The CCG takes $v_c \equiv 0$. The momentum conservation equation of the CDM component leads to $\alpha_{\rm pert} = 0$.

Instead of directly solving equations in this gauge condition, here we use the gauge transformation from the $w$CG to CCG. We consider two coordinate systems: the $w$CG ($x^c$) and the CCG ($\widehat x^c$). Under a gauge transformation $\widehat x^c = x^c + \xi^c$, we have \cite{Bardeen-1988,FNL-HN-2013}
\bea
   & & \widehat \delta_I
       = \delta_I - a {\dot \mu_I \over \mu_I} \xi^0, \quad
       \widehat v_I = v_I - \xi^0,
   \nonumber \\
   & &
       \widehat \kappa = \kappa + \left( 3 \dot H + {\Delta \over a^2}
       \right) a \xi^0, \quad
       \widehat \chi = \chi - a \xi^0,
   \nonumber \\
   & &
       \widehat {\delta \phi}
       = \delta \phi - a \dot \phi \xi^0, \quad
       \widehat \varphi = \varphi - a H \xi^0,
   \label{GT-2}
\eea
where $\xi^c \equiv \xi^c |_{w \rightarrow c}$. We have
\bea
   & & \widehat v_c \equiv 0, \quad
       v_w \equiv 0,
   \label{GT-3}
\eea
as the gauge conditions in the two coordinates. Thus, in the $w$-dominant era and in the large scale limit, we have
\bea
   & & \widehat {\overline v}_w = - a H \xi^0
       = - {\overline v}_c, \quad
       \widehat \delta_w = \delta_w + 3 (1 + w) a H \xi^0,
   \nonumber \\
   & &
       \widehat \delta_c = \delta_c + 3 a H \xi^0, \quad
       \widehat {\overline \kappa} = \overline \kappa
       + 3 a {\dot H \over H} \xi^0, \quad
       \widehat {\overline \chi} = \overline \chi - a H \xi^0,
   \nonumber \\
   & &
       \widehat {\delta \phi} = \delta \phi - a H \phi^\prime \xi^0, \quad
       \widehat \varphi = \varphi - a H \xi^0.
   \label{GT-4}
\eea
Using solutions in the $w$CG in Eqs.\ (\ref{sol-inv-wCG-1}), (\ref{sol-inv-wCG-2}) and (\ref{sol-inv-wCG-3}), and using Eqs.\ (\ref{eq2}) and (\ref{eq3}) we have
\bea
   & &
       \widehat \delta_w = {5 + 3 w \over 5 + 9 w} \delta_w,
   \label{sol-inv-CCG-1}
\eea
and
\bea
   & &
       \widehat \delta_c = {1 \over 1 + w} \widehat \delta_w, \quad
       \widehat {\overline v}_w
       = {2 w \over (1 + w) (5 + 3 w)} \widehat \delta_w,
   \nonumber \\
   & &
       \widehat {\delta \phi} = {3 (1 + 3 w) \over
       (7 + 9 w) \alpha + 29 + 60 w + 27 w^2}
       \phi \widehat \delta_w,
   \nonumber \\
   & &
       {\Delta \over a^2 H^2} \widehat \varphi
       = - {5 + 9 w \over 2 (1 + w)} \widehat \delta_w, \quad
       \widehat {\overline \kappa}
       = {1 + 3 w \over 1 + w} \widehat \delta_w,
   \nonumber \\
   & &
       {\Delta \over a^2 H^2} \widehat {\overline \chi}
       = - {5 + 9 w \over (1 + w) (5 + 3 w)} \widehat \delta_w.
   \label{sol-inv-CCG-2}
\eea
We can show that Eq.\ (\ref{eq5}) is satisfied, and Eq.\ (\ref{eq1}) gives
\bea
   & & \widehat \varphi^\prime
       = - {3 w \over 5 + 3 w} \widehat \delta_w,
   \label{sol-inv-CCG-3}
\eea
which is valid to next order in the large-scale expansion; $\varphi^\prime = 0$ for the mode in Eq.\ (\ref{sol-inv-CCG-2}). We have $\widehat \varphi = \varphi$ and $\widehat \chi = \chi$ to the leading order in the large-scale expansion, but not to the next order.
We can show
\bea
   \widehat \delta_\phi
       =
       { \alpha (-1 + 9w) (1 + 3w)
       \over
       2
       [ (7 + 9w) \alpha
       + 29 + 60 w + 27 w^2 ]} \widehat \delta_w.
   \label{sol-inv-CCG-4}
\eea

Equations  (\ref{sol-inv-CCG-1})-(\ref{sol-inv-CCG-4}) are the complete solutions for inverse power-law potential in the CCG.

\section{Exponential potential}

\subsection{Background solutions}

We consider an exponential potential
\bea
   & & V = V_0 e^{-\lambda \phi},
\eea
with constant $\lambda$ and $V_0$. In the presence of a $w$-fluid, we have a scaling solution with $\mu_\phi \propto \mu_w$, thus $a \propto t^{2 \over 3 (1 + w)}$. In a flat background with a $w$-fluid and the scalar field, we have the solution \cite{Lucchin-Matarrese-1985}
\bea
   & & \mu_\phi = {2 \over 1 - w} V
       = {1 \over 1 + w} \dot \phi^2, \quad
       \phi^\prime = {3(1 + w) \over \lambda},
   \nonumber \\
   & &
       \Omega_\phi = 1 - \Omega_w
       = {24 \pi G ( 1 + w) \over \lambda^2}.
   \label{BG-IC-EXP}
\eea
This solution applies as long as the $w$-fluid and the scalar field dominate the evolution. It is convenient to have
\bea
   & & \phi = - {1 \over \lambda} \ln{\left(
       {1 - w \over 2} {\mu_w \over V_0} {\Omega_\phi \over
       1 - \Omega_\phi}
       \right)}.
   \label{BG-IC-EXP-2}
\eea

\subsection{Solutions in the $w$-fluid comoving gauge ($w$CG)}

In the $w$CG, setting $v_w \equiv 0$, from Eqs.\ (\ref{eq4}), (\ref{eq6-I}), (\ref{eq7-I}) and (\ref{EOM}) we have
\bea
   & & \delta_w^{\prime\prime}
       + {1 - 9 w \over 2} \delta_w^\prime
       + \bigg[ - w {\Delta \over a^2 H^2}
       - {3 \over 2} (1 - w) (1 + 3 w)
   \nonumber \\
   & & \qquad
       + {36 \pi G \over \lambda^2} (1 - w) (1 + w)^2 \bigg] \delta_w
   \nonumber \\
   & & \qquad
       = {12 \pi G \over \lambda} (1 + w)^2
       \left[ 4 \delta \phi^\prime
       + 3 (1 - w) \delta \phi \right],
   \\
   & & \delta \phi^{\prime\prime}
       + {3 \over 2} (1 - w) \delta \phi^\prime
       + \left[ - {\Delta \over a^2 H^2}
       + {9 \over 2} \left( 1 - w^2 \right) \right] \delta \phi
   \nonumber \\
   & & \qquad
       = {3 \over \lambda} \left( 1 - w \right)
       \left( \delta_w^\prime - 3 w \delta_w \right).
\eea
In the large-scale limit, by setting $\delta_w \propto \delta \phi \propto e^{m x}$ we have
\bea
   & & \left[ m - 1 - 3 w \right]
       \left[ m + {3 \over 2} (1 - w) \right]
   \nonumber \\
   & & \quad \times
       \left[ m^2 + {3 \over 2} (1 - w) m
       + {9 \over 2} \left( 1 - w^2 \right) \Omega_w
       \right] = 0,
\eea
thus,
\bea
   & & m = 1 + 3 w, \quad
       - {3 \over 2} (1 - w),
   \nonumber \\
   & & \qquad
       {3 \over 4} (1 - w) \left[ - 1
       \pm \sqrt{ 1 - 8 {1 + w \over 1 - w} \Omega_w } \right].
\eea
For the growing mode we have $m = 1 + 3 w$, and
\bea
   & & \delta \phi
       = {1 \over \lambda} {3 (1 - w) \over 7 + 9 w} \delta_w,
       \quad
       \delta_w \propto a^m.
   \label{sol-exp-wCG-1}
\eea
These solutions were presented in \cite{HN-2001-scaling}.

In the presence of additional but subdominant CDM component the above equations and solutions remain valid. The evolution of CDM component is described additionally by Eqs.\ (\ref{eq6-c-wCG}) and (\ref{eq7-c-wCG}). The growing mode solutions are
\bea
   & & \delta_c = {1 \over 1 + w} \delta_w, \quad
       \overline v_c = - {2 w \over (5 + 9w) (1 + w)} \delta_w,
   \nonumber \\
   & &
       \overline \kappa = {1 \over 1 + w} \delta_w, \quad
       \alpha_{\rm pert} = - {w \over 1 + w} \delta_w,
   \label{sol-exp-wCG-2}
\eea
where we used Eqs.\ (\ref{eq6-I}) and (\ref{eq7-I}).
These solutions are the same as in the inverse power-law case in Eqs.\ (\ref{sol-inv-wCG-2}) and (\ref{sol-inv-wCG-3}). From Eqs.\ (\ref{eq2}) and (\ref{eq3}) we have
\bea
   & & {\Delta \over a^2 H^2} \varphi
       = \left( - {5 + 3 w \over 2 (1 + w)}
       + {9 (5 + 3 w) (1 - w) \over 4 (7 + 9w)} \Omega_\phi \right)
       \delta_w,
   \nonumber \\
   & &
       {\Delta \over a^2 H^2} \overline \chi
       = \left( - {1 \over 1 + w}
       + {9 (1 - w) \over 2 (7 + 9w)} \Omega_\phi \right) \delta_w.
   \label{sol-exp-wCG-3}
\eea
We can show that Eq.\ (\ref{eq5}) is satisfied, and Eq.\ (\ref{eq1}) gives
\bea
   & & \varphi^\prime
       = - \left( {w \over 1 + w}
       + {3 (1 - w) \over 2 (7 + 9w)} \Omega_\phi \right) \delta_w.
   \label{sol-exp-wCG-4}
\eea
This equation is valid to next order in the large-scale expansion; $\varphi^\prime = 0$ for the mode in Eq.\ (\ref{sol-exp-wCG-3}).
We can show
\bea
   \delta_\phi
       =
       { -1 + 24 w + 9w^2
       \over
       2 (7 + 9w)} \delta_w.
   \label{sol-exp-wCG-5}
\eea

Equations (\ref{sol-exp-wCG-1})-(\ref{sol-exp-wCG-5}) are the complete solutions for exponential potential in the $w$CG.

\subsection{Solutions in the CDM comoving gauge (CCG)}

In the CCG, setting $v_c \equiv 0$, from Eq.\ (\ref{eq7-I}) for the CDM component, we have $\alpha_{\rm pert} = 0$.

Using the gauge transformation from the $w$CG to CCG in Eq.\ (\ref{GT-4}), solutions in the $w$CG in Eqs.\ (\ref{sol-exp-wCG-1})-(\ref{sol-exp-wCG-3}) give
\bea
   & &
       \widehat \delta_w = {5 + 3 w \over 5 + 9 w} \delta_w,
   \label{sol-exp-CCG-0}
\eea
and
\bea
   & &
       \widehat \delta_c = {1 \over 1 + w} \widehat \delta_w, \quad
       \widehat {\overline v}_w
       = {2 w \over (1 + w) (5 + 3 w)} \widehat \delta_w,
   \nonumber \\
   & &
       \widehat {\overline \kappa}
        = {1 + 3 w \over 1 + w} \widehat \delta_w, \quad
       \widehat {\delta \phi}
        = {3 \over \lambda} {1 + 3 w
        \over 7 + 9 w} \widehat \delta_w.
   \label{sol-exp-CCG-1}
\eea
We have $\widehat {\varphi} = \varphi$ and $\widehat {\overline \chi} = \overline \chi$ presented in Eq.\ (\ref{sol-exp-wCG-3}), thus
\bea
   & & {\Delta \over a^2 H^2} \widehat \varphi
       = \left( - {5 + 9 w \over 2 (1 + w)}
       + {9 (5 + 9 w) (1 - w) \over 4 (7 + 9w)} \Omega_\phi \right)
       \widehat \delta_w,
   \nonumber \\
   & &
       {\Delta \over a^2 H^2} \widehat {\overline \chi}
       = \left( - {1 \over 1 + w}
       + {9 (1 - w) \over 2 (7 + 9w)} \Omega_\phi \right)
       {5 + 9w \over 5 + 3 w} \widehat \delta_w.
   \label{sol-exp-CCG-2}
\eea
To the next order in the large-scale expansion, from Eq.\ (\ref{sol-exp-wCG-3}) we have
\bea
   & & \widehat \varphi^\prime
       = - \left( {3 w \over 5 + 3w}
       + {3 (1 - w) (5 + 9w) \over 2 (7 + 9w) (5 + 3w)} \Omega_\phi \right) \widehat \delta_w.
\eea
We can show
\bea
   \widehat \delta_\phi
       =
       { (-1 + 9w) (1 + 3 w)
       \over
       2 (7 + 9w)} \widehat \delta_w.
   \label{sol-exp-CCG-4}
\eea

Equations (\ref{sol-exp-CCG-0})-(\ref{sol-exp-CCG-4}) are the complete solutions for exponential potential in the CCG.

\section{Numerical evolution}

The initial conditions presented in this work can be applied to diverse dark energy scenarios based on the MSF as long as these type of potentials approximate the evolution in the early epoch when the initial conditions are imposed. For example, the initial condition for the inverse power-law potential can handle the SUGRA potential \cite{Brax-Martin-1999} with
\bea
   & & V = {{V_0 \over \phi^\alpha} e^{\lambda \phi^2}}.
   \label{SUGRA-potential}
\eea
The initial condition for the exponential potential can handle the double-exponential potential \cite{Park-2009} with
\bea
   & & V = V_1 e^{-\lambda_1 \phi} + V_2 e^{-\lambda_2 \phi},
   \label{double-exp-potential}
\eea
and Albrecht-Scordis (AS) potential \cite{AS-2000} with
\bea
   & & V = V_0 [(\phi -\phi_0)^2 + A ] e^{-\lambda \phi}.
   \label{AS-potential}
\eea

\begin{figure}
\begin{center}
\includegraphics[width=8.9cm]{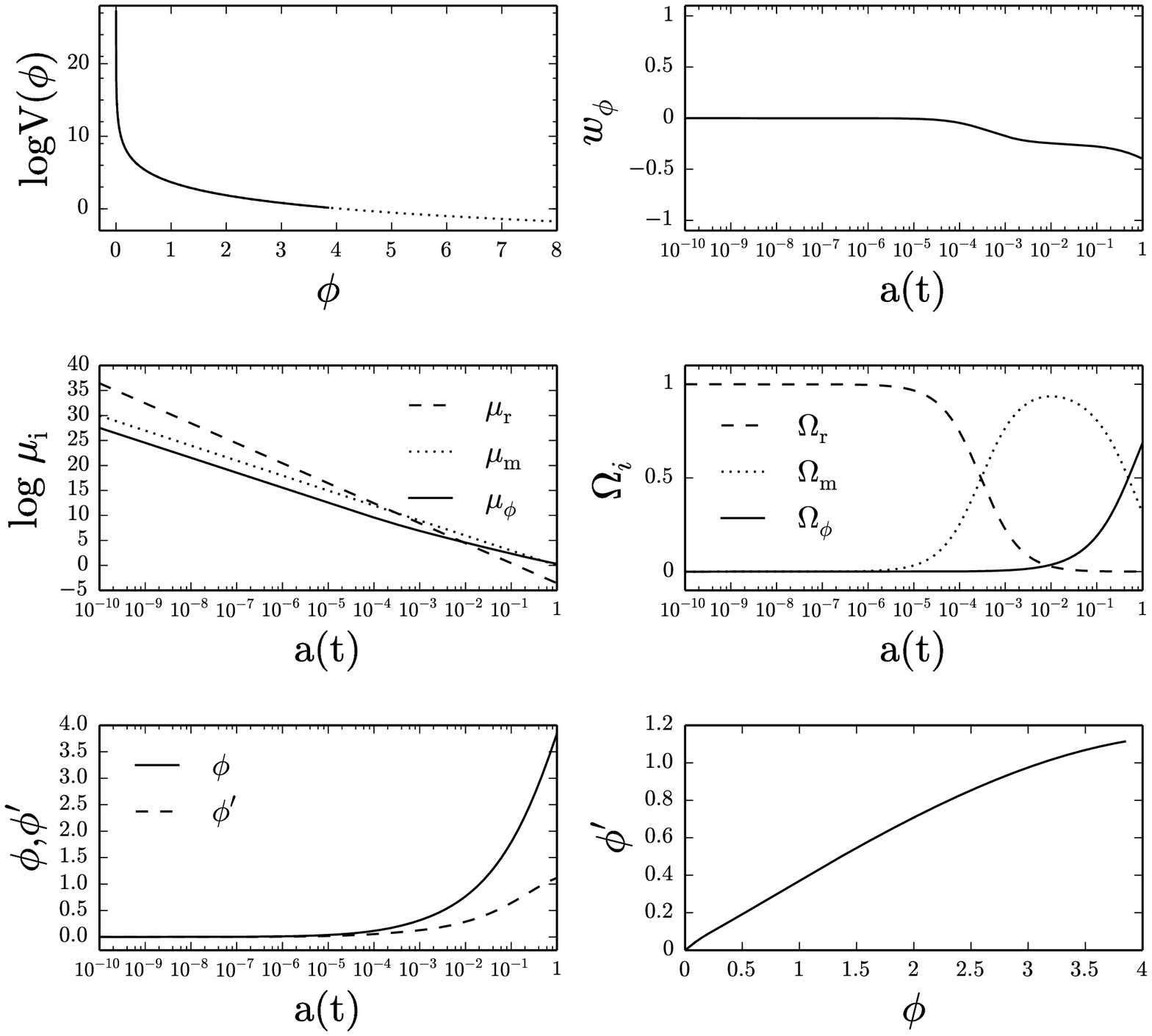}
\includegraphics[width=8.9cm]{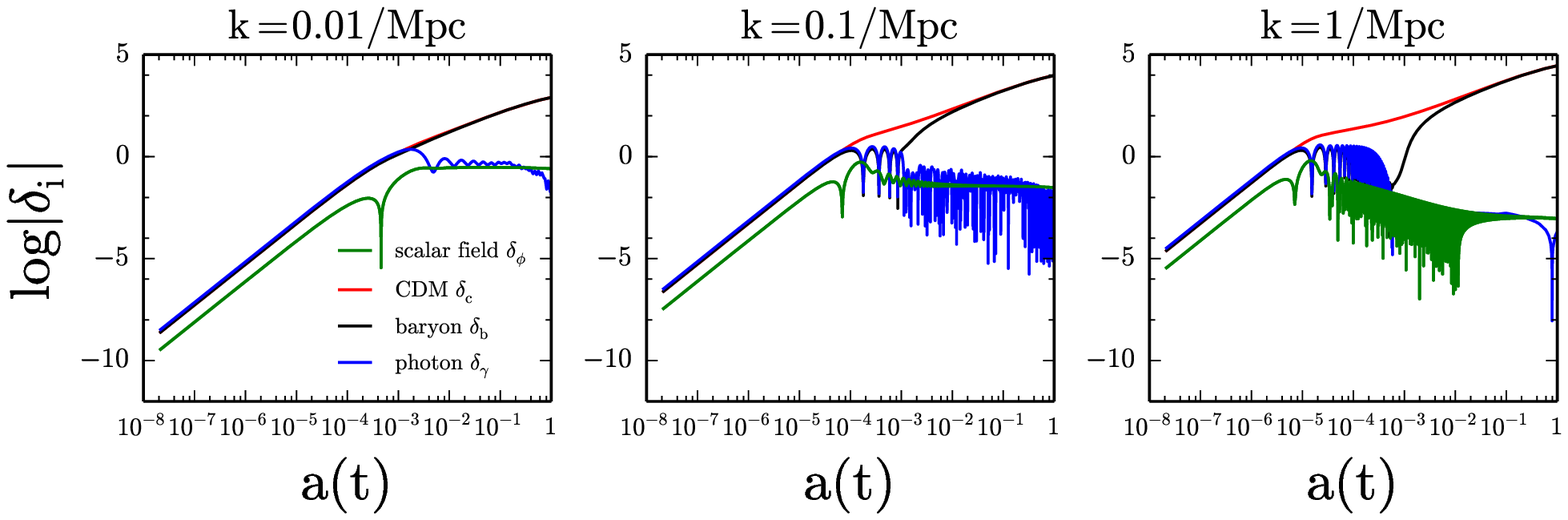}
\caption{
	     Background (top six panels) and perturbation
         (bottom three panels for three different scales)
         evolutions for an inverse power-law potential in Eq.\ (\ref{inv-potential}).
         We take $\alpha = 6$ for the background.
         As the initial conditions we use Eqs.\ (\ref{BG-IC-IPL-1})
         and (\ref{BG-IC-IPL-2}) for the background, and
         Eqs.\ (\ref{sol-inv-CCG-1})-(\ref{sol-inv-CCG-4}) for perturbations in the CCG.
         In the $\phi - V(\phi)$ diagram, the shape of the potential is indicated by the dotted line, and the real evolution till $a_0 = 1$ is indicated by the solid line.
	     }
\label{fig:IPL-potential}
\end{center}
\end{figure}

\begin{figure}
\begin{center}
\includegraphics[width=8.9cm]{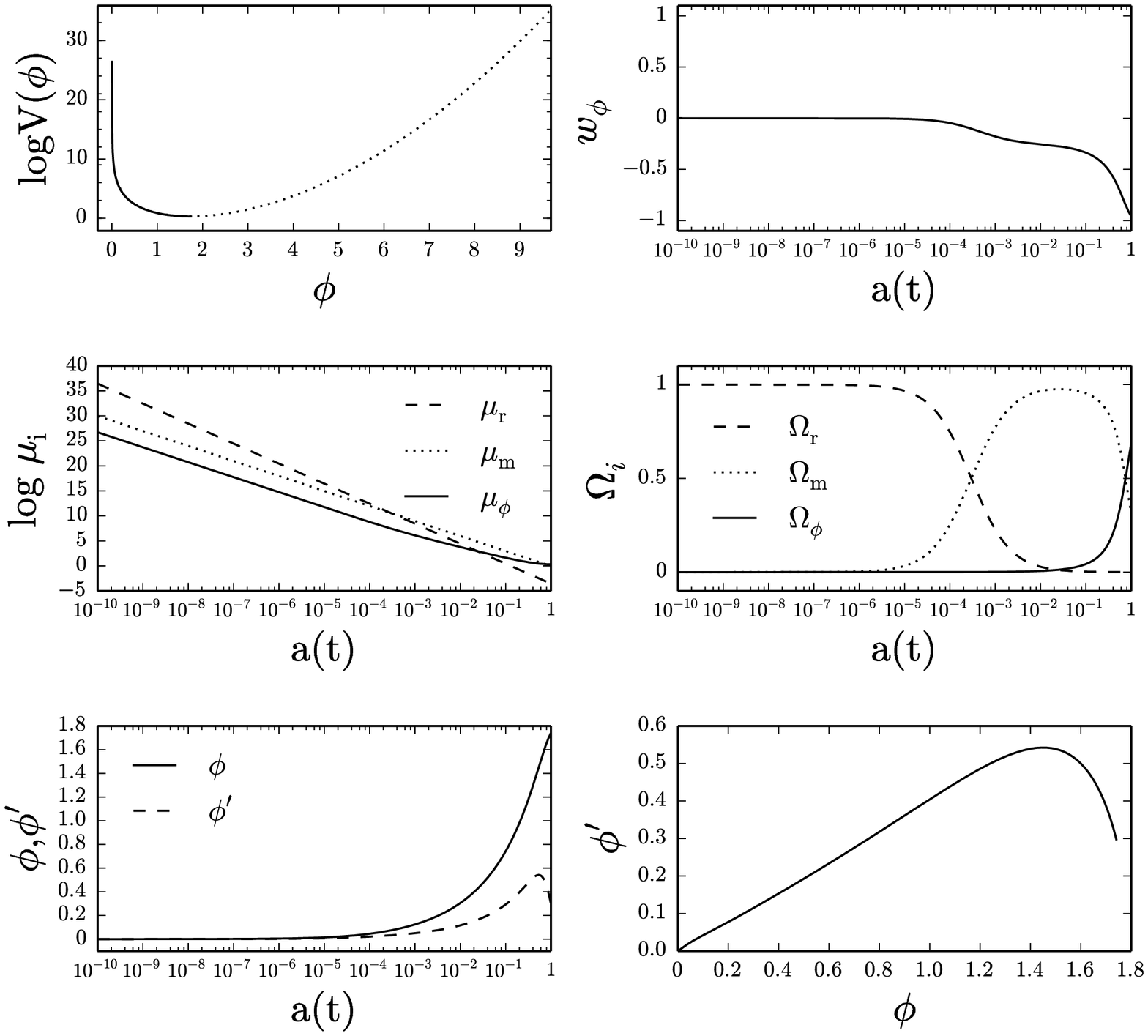}
\includegraphics[width=8.9cm]{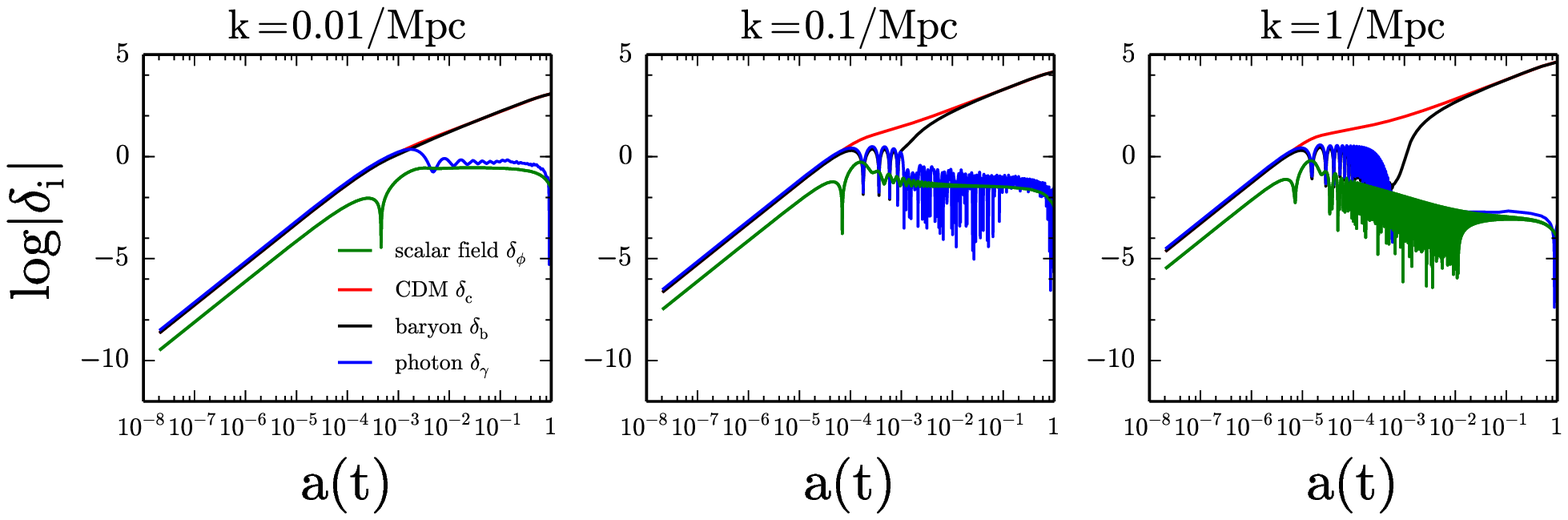}
\caption{
         The same as Fig.\ \ref{fig:IPL-potential}
         for a SUGRA potential in Eq.\ (\ref{SUGRA-potential}).
         We take $\alpha = 6$ and $\lambda = 1$ for the background.
         In this and previous figures, Eq.\ (\ref{sol-inv-CCG-4}) shows that $\widehat \delta_\phi$ changes sign at $w = {1/9}$.
	     }
\label{fig:SUGRA-potential}
\end{center}
\end{figure}

\begin{figure}
\begin{center}
\includegraphics[width=8.9cm]{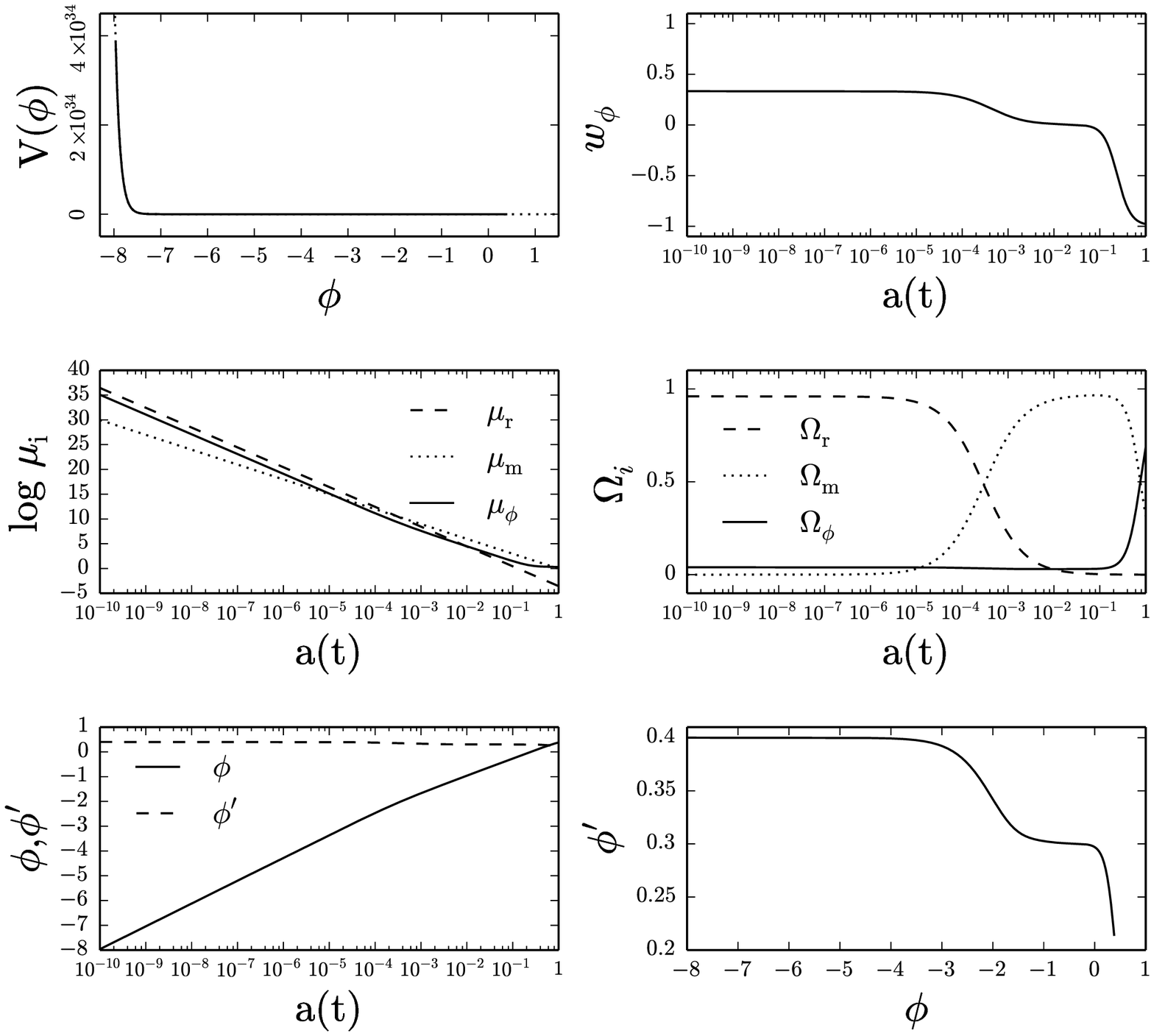}
\includegraphics[width=9.1cm]{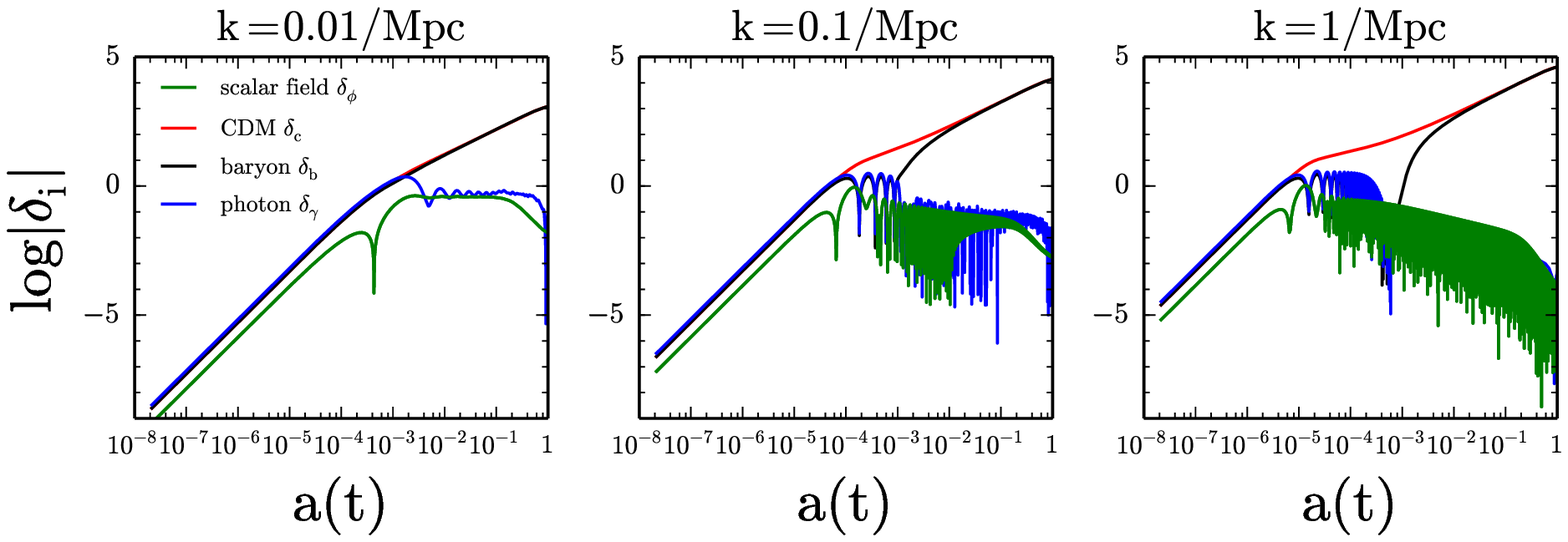}
\caption{
         The same as Fig.\ \ref{fig:IPL-potential}
         for a double exponential potential in Eq.\ (\ref{double-exp-potential}).
         We take $\lambda_1 = 10$, $\lambda_2 = 0.1$ and $V_1 = 1$ for the background.
         In this and next figures, Eq.\ (\ref{sol-exp-CCG-4}) shows that $\widehat \delta_\phi$ changes sign at $w = {1/9}$.
	     }
\label{fig:DE-potential}
\end{center}
\end{figure}

\begin{figure}
\begin{center}
\includegraphics[width=8.9cm]{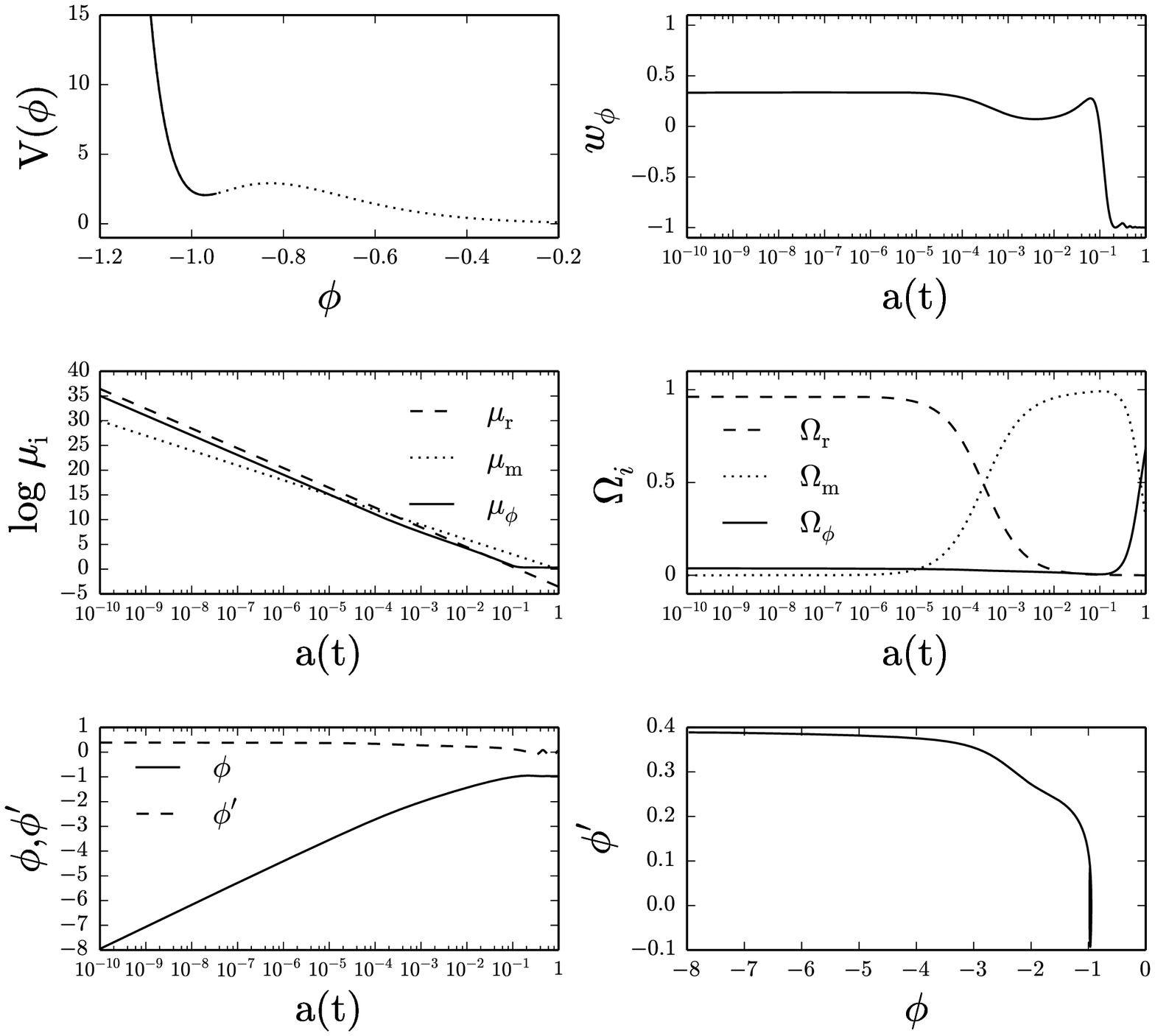}
\includegraphics[width=9.1cm]{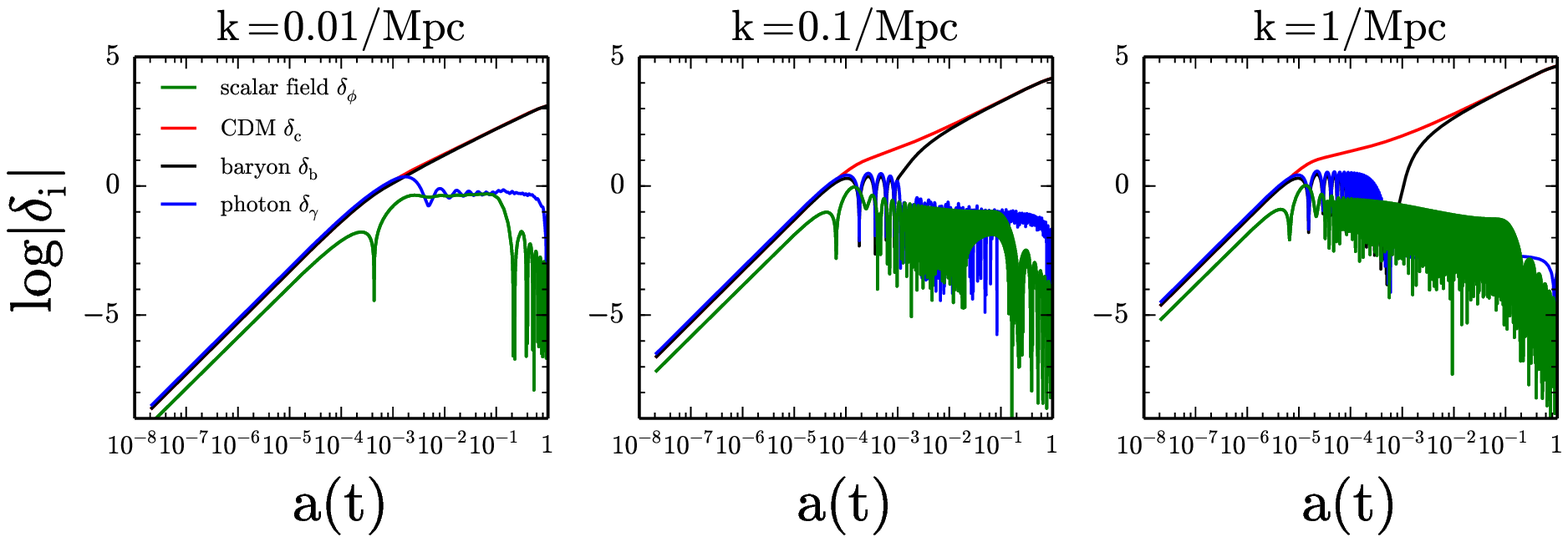}
\caption{
         The same as Fig.\ \ref{fig:IPL-potential}
         for AS potential in Eq.\ (\ref{AS-potential}).
         We take $\lambda_1 = 10$, $\lambda = 10$, $\phi_0 = -1$ and $A = 0.005$ for the background.
         }
\label{fig:AS-potential}
\end{center}
\end{figure}

Evolutions of the background and perturbation variables in the four different models in the CCG are presented in Figures \ref{fig:IPL-potential}-\ref{fig:AS-potential} for the four models in Eqs.\ (\ref{inv-potential}) and (\ref{SUGRA-potential})-(\ref{AS-potential}), respectively. We use the CAMB code \cite{CAMB} modified by supplementing the scalar field for the perturbation as well as background. The perturbations in the CAMB are based on the CCG.

For the background, we take the flat $\Lambda$CDM model parameters constrained by the Planck 2015 CMB data  (TT+lowP) \cite{Planck-2015} without massive neutrino: $h = 0.6731$, $\Omega_b h^2 = 0.02222$, $\Omega_c h^2 = 0.1197$ with the present Hubble constant normalized as $H_0 \equiv 100 h {\rm km}/{\rm sec}/{\rm Mpc}$. In our way of managing the background evolution, we have adjusted one of the potential parameters to satisfy $H(t_0) = H_0$ in the program: these give $\kappa = 4663.7$, $V_0 = 2.7103$, $V_2 = 2.0936$ and $V_0 = 2.1395 \times 10^{-2}$ for the models in Eqs.\ (\ref{inv-potential}) and (\ref{SUGRA-potential})-(\ref{AS-potential}), respectively. Other parameters are explained in the Figure captions.

For perturbations, we take the adiabatic initial condition with amplitudes $\delta_{\gamma, i} = 3.3333 \times 10^{-9}$ for $k = 0.01~\textrm{Mpc}^{-1}$, $\delta_{\gamma, i} = 3.3333 \times 10^{-7}$ for $k = 0.1~\textrm{Mpc}^{-1}$ and $\delta_{\gamma, i} = 3.3333 \times 10^{-5}$ for $k = 1~\textrm{Mpc}^{-1}$ at $a_i = 2.1575 \times 10^{-8}$. The massless neutrino has the same amplitude as the photon, and the baryon and CDM have amplitude adiabatically related to the photon, thus $\delta_b = \delta_c = {3 \over 4} \delta_\gamma$ [In our numerical work, as we consider the CCG only we ignore hats on perturbation variables.]

\begin{figure}
\begin{center}
\includegraphics[width=8.9cm]{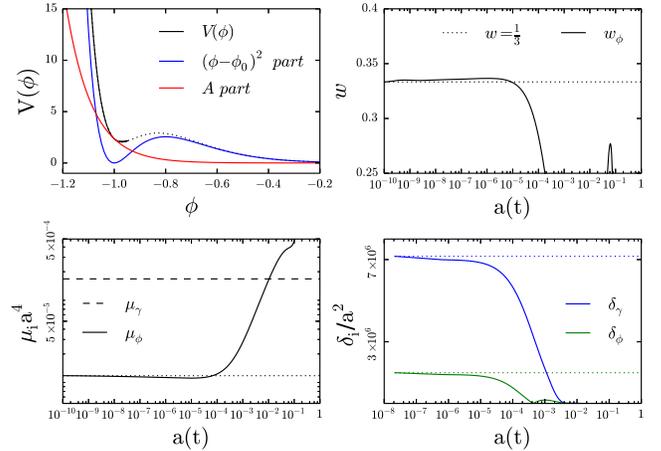}
\caption{The same model as in Fig.\ \ref{fig:AS-potential}, now
         showing the approximate nature of our solutions in the AS model. Notice the small difference of the $w_\phi$, $\mu_\phi$, $\delta_\phi$ and $\delta_\gamma$ compared with the dotted line (the scaling solution) in the early era (till around $a \sim 10^{-5}$). Except for this understandable difference in details our solution is still quite successful.
         }
\label{fig:AS-potential-detail}
\end{center}
\end{figure}

Figures show that our initial conditions for the background and perturbations in the CCG work well. The model used in Fig.\ \ref{fig:AS-potential} deserves a special notice. For the AS potential, in the early era when the initial condition is imposed, $(\phi -\phi_0)^2 e^{-\lambda \phi}$ part dominates whereas our initial conditions were derived for pure $e^{-\lambda \phi}$. However, as the exponential part dominates over the quadratic part in the early epoch, our solution is still valid approximately. In Fig.\ \ref{fig:AS-potential} we cannot distinguish the difference by eye. In order to show the approximate nature and the small difference in detail, we separately present the potential and evolutions of background and perturbations in Fig.\ \ref{fig:AS-potential-detail}.

\section{Discussion}

We presented the perturbation initial conditions for a fluids-field system with two types of scalar field potential in two gauge conditions. The fluids-field system includes an ideal fluid with constant $w$, a minimally coupled scalar field, and a subdominant zero-pressure fluid. The solutions are: for inverse power-law potential, Eqs.\ (\ref{sol-inv-wCG-1}), (\ref{sol-inv-wCG-2})-(\ref{sol-inv-wCG-5}) in the $w$CG and Eqs.\ (\ref{sol-inv-CCG-1})-(\ref{sol-inv-CCG-4}) in the CCG, and for exponential potential, Eqs.\ (\ref{sol-exp-wCG-1})-(\ref{sol-exp-wCG-5}) in the $w$CG and Eqs.\ (\ref{sol-exp-CCG-0})-(\ref{sol-exp-CCG-4}) in the CCG.

As we have shown in the previous section the CCG initial conditions for the two type of potentials will have wide applications in the Markov chain Monte Carlo parameter estimation with the scalar field as the dark energy. We have implemented the scalar field in the CAMB \cite{CAMB}. The CAMB is based on the synchronous gauge with additional fixing of the remnant gauge mode by setting $v_c \equiv 0$, and this is the same as the CCG. Thus our perturbation solutions in the CCG are suitable for initial conditions for CAMB supplemented by a scalar field as the dark energy. We will study the role of scalar field as the dark energy in future.

\vskip .5cm
%
%
J.H.\ was supported by Basic Science Research Program through the National Research Foundation (NRF) of Korea funded by the Ministry of Science, ICT and future Planning (No.\ 2016R1A2B4007964 and No.\ 2018R1A6A1A06024970). H.N.\ was supported by National Research Foundation of Korea funded by the Korean Government (No.\ 2018R1A2B6002466). C.-G.P. was supported by the Basic Science Research Program through the National Research Foundation of Korea (NRF) funded by the Ministry of Education (No. 2017R1D1A1B03028384).

\section*{Appendix: Basic equations}

We present a complete set of background and scalar-type perturbation equations of multiple component fluids and a scalar field system in a flat Friedmann world model. Our metric and energy-momentum tensor convention is
\bea
   & & d s^2 = - a^2 \left( 1 + 2 \alpha \right) d \eta^2
       - 2 a \chi_{,i} d \eta d x^i
   \nonumber \\
   & & \qquad
       + a^2 \left( 1 + 2 \varphi \right) \delta_{ij} d x^i d x^j,
   \\
   & & T^0_0 = - \left( \mu + \delta \mu \right), \quad
       T^0_i = - \left( \mu + p \right) v_{,i},
   \nonumber \\
   & &
       T^i_j = \left( p + \delta p \right) \delta^i_j.
\eea
Background equations are
\bea
   & & H^2 = {8 \pi G \over 3} \mu
       + {\Lambda \over 3},
   \\
   & & \dot \mu + 3 H \left( \mu + p \right)
       = 0,
\eea
where $H \equiv \dot a/a$ with an overdot indicating a time derivative based on $t$ with $dt \equiv a d \eta$. In the multiple component case, for individual component, we have
\bea
   & &
       \dot \mu_I + 3 H \left( \mu_I + p_I \right)
       = 0,
\eea
and
\bea
   & & \mu = \sum_J \mu_J, \quad
       p = \sum_J p_J.
\eea
For a MSF we have
\bea
   & & \ddot \phi + 3 H \dot \phi + V_{,\phi} = 0,
   \\
   & & \mu_\phi = {1 \over 2} \dot \phi^2 + V, \quad
       p_\phi = {1 \over 2} \dot \phi^2 - V.
\eea

Ignoring the anisotropic stress, the scalar-type perturbation equations without taking the temporal gauge (slicing) condition are \cite{Bardeen-1988,HN-2001-scaling}
\bea
   & & \kappa = 3 H \alpha - 3 \dot \varphi
       - {\Delta \over a^2} \chi,
   \label{eq1} \\
   & & 4 \pi G \delta \mu
       + H \kappa
       + {\Delta \over a^2} \varphi = 0,
   \label{eq2} \\
   & & \kappa + {\Delta \over a^2} \chi
       - 12 \pi G a \left( \mu + p \right) v = 0,
   \label{eq3} \\
   & & \dot \kappa + 2 H \kappa
       + \left( 3 \dot H + {\Delta \over a^2} \right) \alpha
       = 4 \pi G \left( \delta \mu + 3 \delta p \right),
   \label{eq4} \\
   & & \varphi + \alpha
       - \dot \chi - H \chi
       = 0,
   \label{eq5} \\
   & & \delta \dot \mu
       + 3 H \left( \delta \mu + \delta p \right)
       + \left( \mu + p \right) \left( 3 H \alpha
       - \kappa - {\Delta \over a} v \right)
   \nonumber \\
   & & \qquad
       = 0,
   \label{eq6} \\
   & & {1 \over a^4} \left[ a^4 \left( \mu + p \right)
       v \right]^{\displaystyle\cdot}
       = {1 \over a} \left[ \delta p
       + \left( \mu + p \right) \alpha \right].
   \label{eq7}
\eea
For each component, the conservation equations are
\bea
   & & \delta \dot \mu_I
       + 3 H \left( \delta \mu_I + \delta p_I \right)
   \nonumber \\
   & & \qquad
       + \left( \mu_I + p_I \right) \left( 3 H \alpha
       - \kappa - {\Delta \over a} v_I \right)
       = 0,
   \label{eq6-I} \\
   & & {1 \over a^4} \left[ a^4 \left( \mu_I + p_I \right)
       v_I \right]^{\displaystyle\cdot}
       = {1 \over a} \left[ \delta p_I
       + \left( \mu_I + p_I \right) \alpha \right],
   \label{eq7-I}
\eea
and we have
\bea
   & &
       \delta \mu = \sum_J \delta \mu_J, \quad
       \delta p = \sum_J \delta p_J,
   \nonumber \\
   & &
       v = {\sum_J ( \mu_J + p_J ) v_J \over
       \sum_K (\mu_K + p_K)}.
\eea
For the scalar fields we have
\bea
   & & \delta \ddot \phi
       + 3 H \delta \dot \phi
       - {\Delta \over a^2} \delta \phi
       + V_{,\phi\phi} \delta \phi
   \nonumber \\
   & & \qquad
       = 2 \ddot \phi \alpha
       + \dot \phi \left( \dot \alpha
       + 3 H \alpha + \kappa \right),
   \label{EOM}
\eea
and
\bea
   & &
       \delta \mu_\phi
       = \dot \phi \delta \dot \phi
       - \dot \phi^2 \alpha
       + V_{,\phi} \delta \phi,
   \nonumber \\
   & &
       \delta p_\phi
       = \dot \phi \delta \dot \phi
       - \dot \phi^2 \alpha
       - V_{,\phi} \delta \phi, \quad
       v_\phi = {1 \over a}
       {\delta \phi
       \over \dot \phi}.
\eea

%
%


\end{document}